\documentclass[twocolumn]{aastex63}

\received{2025 September 23}
\revised{2025 October 27}
\revised{2025 November 19}
\accepted{2025 November 20}

\usepackage{enumitem} 
\usepackage{lipsum}
\usepackage{amsmath}
\usepackage{booktabs}

\usepackage{lineno}
\usepackage{afterpage}
\shortauthors{Blackwell et al.}
\graphicspath{{./}{figures/}}

\begin{document}

\defcitealias{Blackwell2025}{B25}

\title{Tracing Early Cosmic Chemical Enrichment: A Uniform XMM-Newton Survey of Metallicity in Galaxy Groups and Clusters}

\author[0000-0002-8195-0563]{Anne E. Blackwell}
\affiliation{University of Michigan, Ann Arbor, MI 48104, USA}

\author[0000-0001-6276-9526]{Joel N. Bregman}
\affiliation{University of Michigan, Ann Arbor, MI 48104, USA}

\author[0009-0000-8193-7855]{Sophia Chan Davis}
\affiliation{University of Texas at San Antonio, San Antonio, TX 78249, USA}











\begin{abstract}

Observed metal abundances in the intracluster medium (ICM) of galaxy groups and clusters, $Z_{ICM}$, exceed what is expected from present-day stellar populations alone. 
Galaxy clusters are presumed to be near closed-box systems, allowing constraints to be placed on the origins of metals and stellar populations responsible for $Z_{ICM}$. 
We present a uniform XMM-Newton survey of 26 galaxy groups and clusters, measuring radial metallicity profiles and relating $Z_{ICM}$ with the stellar fraction $M_*/M_{gas}$. 
We determine $Z_{ICM}$ via spectral fitting across multiple annuli finding a best fit of $Z_{ICM} = -0.08^{+0.07}_{-0.07}\, log\left(\frac{M_*}{M_{gas}}\right) + 0.30^{+0.06}_{-0.06}$ with intrinsic scatter $\sigma_p = 0.09^{+0.02}_{-0.01}$. 
We use closed-box chemical evolution models to estimate the metallicity yield from observable stellar populations, incorporating updated supernova yields and corrections for metals locked in remnants, $Z_* = (1.14 \pm 0.52) \, log\left(1 + \frac{M_*}{M_{gas}}\right)$. 
Our results demonstrate that present-day stellar populations systematically underpredict $Z_{ICM}$, with an inferred excess component increasing in systems with low $M_*/M_{gas}$. 
This trend supports the need for an early enrichment population (EEP) distinct from visible stars, $Z_{EEP}$. 
We find this necessity holds when reconsidering the closed-box assumption by removing all galaxy groups, potential leaky systems, deriving $Z_{EEP}$ within $1\sigma$ when including and excluding groups. 
Three systems (NGC1132, NGC5098, and NGC4325) deviate from the survey trend, exhibiting steep negative radial metallicity gradients and unusually low $Z_{ICM}$. 
We posit these systems to be late-forming whose ICM enrichment reflects only recent stellar populations. 
Our analysis quantifies the necessity of an EEP and provides trends for testing cluster chemical evolution models.

\end{abstract}


\section{Introduction} \label{sec:intro}

Metals serve as vital tracers of stellar evolution, feedback processes, and the integrated history of star formation throughout the Universe. 
In particular, the intracluster medium (ICM) of galaxy clusters is a critical environment for studying cosmic chemical enrichment \citep[e.g.,][]{Baldi2012, Mantz2017, Mernier2017, Ghiz2020, Lovisari2020, Gatuzz2023, 2023bGatuzz}. 
Galaxy clusters possess deep potential wells that retain enriched material, making them ideal laboratories for tracing the production of metals over cosmic time.

Chemical enrichment models often adopt a ``closed-box'' framework, in which gas and metals are neither lost nor gained from the system. 
The observed ICM metallicity, $Z_{ICM}$, should be fully accounted for by the metals produced by the current observable stellar populations, $Z_*$, under an assumed initial mass function (IMF) within this model. 
However, a growing body of evidence suggests that this assumption may not adequately describe the observed metallicity trends, particularly in systems with low stellar mass fractions, $M_*/M_{gas}$ \citep{Breg2010, Blackwell2022, Blackwell2025}. 
Prior works indicate that closed-box models systematically underpredict $Z_{ICM}$, implying that an additional stellar population may have contributed significantly to an early enrichment, the early enrichment population (EEP) \citep{Elbaz1995, Loew2001, Breg2010, Loew2013, Renzini2014, Blackwell2022, Blackwell2025}.

Several alternative theories have been proposed to resolve this discrepancy. 
\cite{Zibetti2005} suggested that the missing stellar population may be found in diffuse intracluster light (ICL), but subsequent observations showed that the ICL does not contain enough mass to account for excess metals. 
\cite{Renzini2014} provide a comprehensive review of the broader metal budget problem and outline various scenarios that could explain the enrichment, including hierarchical assembly and selective metal loss. 
More recently, \cite{Biffi2025} used cosmological simulations to model the origin of metals in galaxy groups and clusters. 
They find that the stellar populations within their simulations accurately recreate $Z_{ICM}$ and therefore claim to alleviate the iron budget tension.
However, they overpredict the stellar mass and consequently the iron locked in stars, therefore underpredicting the iron share, the ratio of Fe diffused in the ICM to that locked in stars, \citep{Renzini2014, Ghiz2020} by approximately an order of magnitude compared to observations.
Despite these efforts and tensions, the challenge first identified by \cite{Elbaz1995} remains unresolved, and the EEP hypothesis has not yet been conclusively ruled out.

Previous efforts to constrain the EEP include the works of \cite{Blackwell2022} and \cite{Blackwell2025}, who analyzed trends in $Z_{ICM}$ as a function of $M_*/M_{gas}$. 
In particular, \citet{Blackwell2025}, hereafter \citetalias{Blackwell2025}, conducted a uniform analysis of archival XMM-Newton data for 26 galaxy groups and clusters with $M_{*,500}$ and $M_{gas,500}$ from \citet{Lagana2013}. 
They derived plausible IMFs for the EEP that could account for the observed deviations from closed-box predictions.

In this study, we present the full dataset and X-ray spectral fit results for the 26 galaxy groups and clusters comprising the survey. 
We evaluate the trend of $Z_{ICM}$ vs $M_*/M_{gas}$ under a closed-box chemical evolution model and demonstrate the need for an early enrichment component. 
We derive the expected contribution from present-day stellar populations, $Z_*$, using updated yields from Type Ia and core-collapse supernovae and accounting for metals locked in stellar remnants. 
This methodology builds upon that introduced by \citet{Loew2013}, incorporating more recent estimates of supernova rates, IMF variations, and additional consideration of metals locked in remnants.

The residual metallicity, $Z_{EEP} = Z_{ICM} - Z_*$, is the necessary contribution from the theorized EEP required to reproduce observations of $Z_{ICM}$.
The contribution of $Z_{EEP}$ may vary across systems as it may be entangled with other cluster properties such as entropy profiles, cuspiness, or dynamical history. 
We investigate the dependence of $Z_{EEP}$ on these properties in a companion study (Gherri et al., in prep).

For this paper, we assume a $\Lambda$CDM cosmology with $H_0 = 70$ km s$^{-1}$ Mpc$^{-1}$, a baryon density parameter of $\Omega_b = 0.046$, dark matter density parameter of $\Omega_{DM} = 0.24$, and the total baryon density parameter is $\Omega_m = 0.29$ \citep{Planck2018cosmo}.
We also adopt a diet-Salpeter IMF as developed by \cite{Bell2003} and adopted in \cite{Blackwell2022} and \citetalias{Blackwell2025}.
This IMF has a slope of $\alpha = 0.35$ in the mass range $0.5-1 M_\odot$, and a slope of $\alpha = 1.35$ in the mass range $1-150 M_\odot$ for an IMF defined as $\phi(m)= m^{-(1+\alpha)}$ (per unit log mass).

This paper is organized as follows. 
In Section~\ref{sec:data}, we describe the survey selection, data reduction, and spectral analysis methodology. 
Section~\ref{sec:fitting} presents the fit to the observed $Z_{ICM}$ vs $M_*/M_{gas}$ relation and the derivation of $Z_*$ and $Z_{EEP}$. 
In Section~\ref{sec:discussion}, we discuss the implications of our findings, including assumptions, systematics, and the role of outlier systems. 
We conclude in Section~\ref{sec:conclusions}.

\section{Survey Data Reduction and Analysis} 
\label{sec:data}

Our sample of 26 galaxy groups and clusters follows from  \citetalias{Blackwell2025}, a subset of galaxy groups and clusters identified in \cite{Lagana2013}.
Each system has uniformly derived $M_{*,500}$ and $M_{gas,500}$, using the critical overdensity, and sufficient archival XMM Newton data to determine $Z_{ICM}$ to approximately 10\% precision.

We adopt $R_{500}$ values from \cite{Lagana2013}, which were derived for galaxy groups in that work and for galaxy clusters in \cite{Maughan2008}.
In \cite{Maughan2008}, $R_{500}$ was determined iteratively using the $Y_X - M_{500}$ scaling relation determined by \cite{Vikhlinin2006} \citep{Kravtsov2006}.
The iterative process involved measuring the plasma temperature and the gas mass within the region $0.15-1 R_{500}$ until $R_{500}$ converged within 1\%.
For galaxy groups, \cite{Lagana2013} computed $R_{500}$ using their Equation 6 from \cite{LimaNeto2003}, assuming an isothermal temperature profile measured within 300 kpc.

Our sample inherits the selection criteria of \cite{Lagana2013} who combined nine galaxy groups and 114 clusters spanning $0.02 < z < 1.3$ and $M_{500} \sim 10^{13} - 4 \times 10^{15} M_\odot$.
Of those, 36 systems had sufficient optical and X-ray data to determine stellar and gas masses out to $R_{500}$.
This sample is X-ray luminosity limited to $L_X = 1.9\times 10^{43} \text{erg/s}$.
This criterion inherently favors X-ray bright, gas-rich systems, excluding low-mass or gas-poor groups with insufficient X-ray observations.
That is, the systems we chose for this study favor high signal-to-noise observations.
Prioritizing observation quality ensures reliable measurements of $Z_{ICM}$, $M_*$ and $M_{gas}$, but at the cost of reduced sample completeness.
Our sample selection emphasizes measurement precision over full representation of the $M_*/M_{gas}$ range.
These excluded systems would likely populate the high end of the $M_*/M_{gas}$ distribution, potentially biasing our sample.
We do not explicitly correct for this sample incompleteness within our modeling. 
The derived trends therefore should only be interpreted for X-ray luminous systems within $0.039 < M_*/M_{gas} < 0.48$.

The final sample consideration is the contribution from non-thermal pressure.
Values adopted from \cite{Maughan2008} and \cite{Lagana2013} either do not, or very briefly touch on the contribution from non-thermal pressure, and only in the galaxy group regime.
Simulations find that non-thermal pressure may contribute $\sim10\%$ of the total pressure support for relaxed galaxy clusters \citep{Lau2009}.
This is a very small contribution, and likely encapsulated in the errors of $M_*$ and $M_{gas}$ as the errors are $\sim30\%$ and $\sim7\%$ on average, respectively.
We do not consider any corrections from non-thermal pressure in this survey.

\subsection{XMM-Newton Data Reduction and Analysis} 
\label{sec:xmm}

We followed the same data reduction and analysis outlined in \cite{Blackwell2022} which follow from \cite{Snowden2008}, using the XMM-ESAS Cookbook, and the Extended Source Analysis Software.
We used SAS version (19.1.0)\footnote{https://www.cosmos.esa.int/web/xmm-newton/sas}, and CALDB (version 4.9.3)\footnote{http://cxc.harvard.edu/caldb/}.
We examined each light curve to ensure proper exclusion of flare intervals, and used \textit{emanom} to check for anomalous chips which were then excluded from the spectral extractions. 

We extracted spectra from five radial bins defined as $<0.1 R_{500}$, $0.1-0.21 R_{500}$, $0.21-0.39 R_{500}$, $0.39-0.53 R_{500}$, and $0.53-0.7 R_{500}$.
These are the same regions as \citetalias{Blackwell2025}.
The value $R_{500}$ is defined as the radius at which the average cluster density is 500 times that of the Universe at the cluster redshift. 

We accounted for two known chip phenomena during spectral extraction: the point-spread function and PN-chip counting of X-ray photons. 
The point-spread function was considered by generating crossarfs which redistribute the X-ray photons into the most probable annuli they physically originated. 
We used \textit{evselect} and \textit{arfgen} to generate additional ARF files to account for the crossarf.
For the PN chip, we extracted two spectra using PATTERN=0 for a low-energy spectrum (0.6-2 keV), and PATTERN=4 for a high-energy spectrum (1.0-7.0 keV). 

\begin{figure}
    \centering
    \includegraphics[width=1\linewidth]{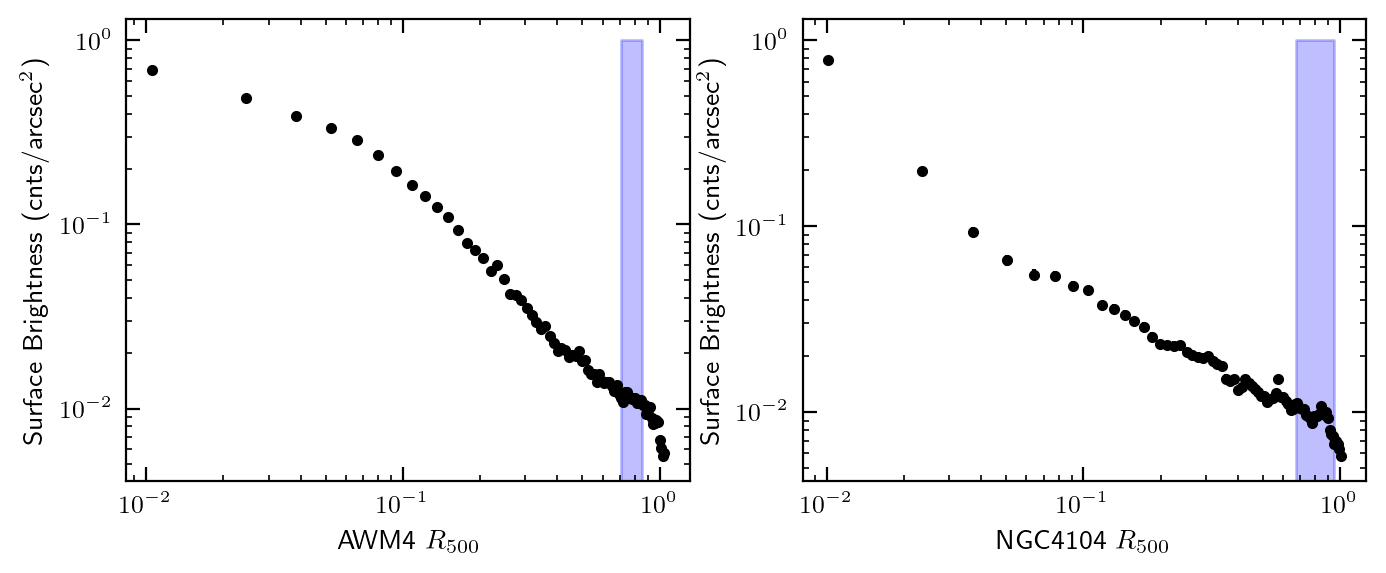}
    \caption{Surface brightness profiles of AWM4 (left) and NGC4104 (right). The blue highlighted region shows where the background region was defined.}
    \label{fig:rad_prof}
\end{figure}

On-chip backgrounds were determined outside of $R_{500}$ for 23 of the systems. 
NGC4104 and AWM4 have $R_{500}$ values of 14.8$\arcmin$ and 14.1$\arcmin$, respectively.  
That means the cluster takes up nearly the entire FOV of the XMM Newton MOS and PN chips, making it impossible to get an on-chip background outside of $R_{500}$.
For these clusters, we took background regions that were as far away from the center of the cluster as possible. 
These backgrounds were selected based on the radial surface brightness profiles, shown in Figure \ref{fig:rad_prof}, as the shaded regions where cluster brightness is negligible.
If there was excess cluster signal in the chosen background, it would get modeled out at part of the background decreasing the fit $Z_{ICM}$.
However, that does not seem to be the case as both clusters have fit $Z_{ICM}$ higher than, or approximately the mean $Z_{ICM}$ at their respective $M_*/M_{gas}$.

We first fit the background spectra using XSPEC (version 12.12.1c)\footnote{https://heasarc.gsfc.nasa.gov/docs/xanadu/xspec/issues/archive/issues.12.12.1c.html} to determine the background contribution.
We modeled the background emission as two gaussian models to capture the XMM Newton instrumental lines, and three \textit{apec} models for the Milky Way Local Hot Bubble and the warm- and hot-Milky Way halo components.
For the former we assume solar abundance and an initial temperature of 0.1 keV, and for the latter we assume an abundance of $0.3 Z_\odot$ and initial temperatures of 0.3 keV and 0.7 keV, respectively.
We multiply the warm- and hot-Milky Way halo component by a \textit{tbabs} model to account column absorption. 
The plasma temperatures are allowed to vary but consistently checked to ensure they are physical. 
A broken powerlaw with index 1.46 is also included for any residual soft proton background.
We assume abundances from \cite{Asplund2009}.
These model components were frozen while fitting for cluster emission.
This methodology ensures that our background is purely background as any overlap of the background region with group or cluster emission would result in poor fit statistics.

We add an additional \textit{apec} and \textit{tbabs} for the cluster component. 
Constants are included to account for differences in solid angle so that spectra from neighboring annuli may be fit simultaneously, allowing for better consideration of the PSF.
Finally, we include an additional \textit{broken-powerlaw} and absorbed \textit{apec} model for the quiescent particle background and to link the models for the respective crossarfs.

\subsection{Chandra Data}
We chose not to include Chandra observations in this study as Chandra's smaller effective collecting area and lower spectral resolution make it less optimal for our analysis objectives.
Due to the reduced collecting area, we are unable to obtain on-chip background measurements for all clusters in our sample.
The lower spectral sensitivity further limits our ability to obtain reliable spectral fits in the outer annuli.
We found that for 75\% of the clusters only one or two annuli contributed to $Z_{ICM}$ since not all three outer annuli could be extracted or fit.
This restriction could bias $Z_{ICM}$ toward higher values, given the typical radial decrease in Fe and our inability to probe larger radii.

We compared $Z_{ICM}$ measurements with and without Chandra observations to ensure the validity of this decision. 
All data analysis and reduction was done using standard reduction practices, as outlined by the extended source analysis guide\footnote{https://cxc.cfa.harvard.edu/ciao/guides/esa.html}.
Including measurements from Chandra altered $Z_{ICM}$ by 4\% on average.
This modest change stems from the larger uncertainties on metallicity values derived from Chandra, compared to those obtained with XMM Newton.
The larger errors result in lower weights on the calculation of $Z_{ICM}$ and our inability to include the outer annuli due to the chip size restrictions of Chandra or insufficient photon statistics. 
For these reasons, we have chosen not to incorporate archival Chandra data in our analysis.

\subsection{Anomalous Systems}

We identified an additional two clusters in \cite{Lagana2013} that met the criteria, however, in a further study, neither system is suitable for the study as we limit this study to virialized systems.
Galaxy cluster A115 is a merging cluster with two asymmetric X-ray peaks and a radio feature \citep[e.g.,][]{Barrena2007, Kim2019}.
We cannot assume azimuthal symmetry for the X-ray emission of this cluster because of the merging X-ray halos and therefore exclude this cluster for our survey.

The second group is NGC5098. 
This system did not initially present as unusual in its X-ray characteristics, however the determined $Z_{ICM} = 0.150 \pm 0.007$ is over 7$\sigma$ away from  NGC4325, the next-lowest $Z_{ICM}$, with $Z_{ICM}$ of $0.205 \pm 0.016$.
We have chosen to exclude NGC5098 from the survey and any subsequent analysis due to its unusually low $Z_{ICM}$.

\subsection{Survey Fit Results}

The fit results are described in Tables \ref{tab:kT_summary} and \ref{tab:z_summary} for the kT and $Z_{ICM}$ fit values, respectively. 

A couple of the observations lacked signal to obtain fit values in the outermost bin.
Those bins are left blank in the tables. 
Other bins are defined by a single value without an error, due to nonphysical kT values being fit otherwise. 
For those fits, we froze the plasma temperature to a single value in order to obtain a reasonable fit, reduced $\chi^2 < 2$.
The frozen plasma temperature was either determined by neighboring bins, or fit results from other observations of the same cluster in the same bin.
We ensured that the frozen kT value did not significantly alter the resulting fit of the metallicity and therefore not arbitrarily determining $Z_{ICM}$.

\begin{figure}
    \centering
    \includegraphics[width=1.0\linewidth]{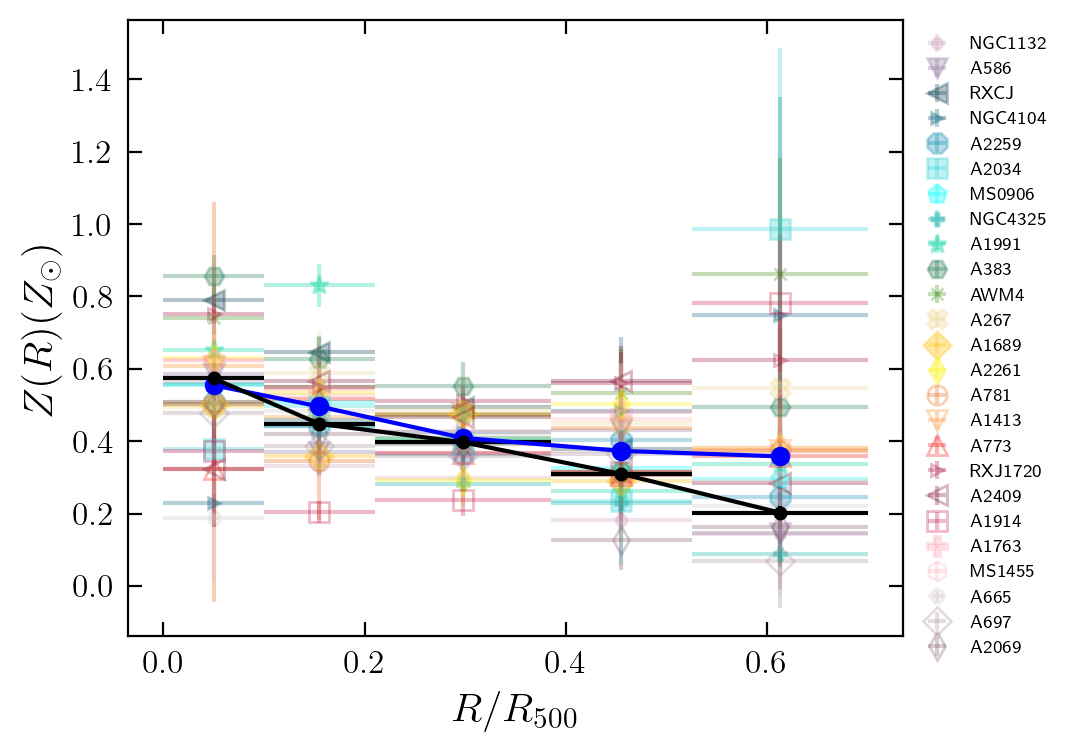}
    \caption{Radial trend of the fit metallicity for all clusters. We show the weighted average (black points) and median (blue points) within each bin. The median trend flattens out to a mean metallicity of $0.37 Z_\odot$ in the outer radial bins. The weighted average trend shows a negative trend of $Z_{ICM}$ with radius best described by a line with slope $-0.70 \pm 0.04$ and intercept $0.57 \pm 0.01$.}
    \label{fig:z_radial}
\end{figure}

In Figure \ref{fig:z_radial} we show the fit metallicity radial trend of all the clusters within our sample (black line), $Z(R)$ (further discussed in Section \ref{sec:outliers}).
We determine the average $Z(R)$ as the weighted average (black data points and trend) and the median (blue data points and trend).
The weighted average trend shows a decreasing $Z(R)$, consistent with \cite{Mernier2017} and simulations \citep{Rasia2015, Dylan2024}.
However, the weighted average emphasizes clusters with smaller uncertainties, potentially biasing the measurement to brighter or more nearby clusters.
The median trend is less susceptible to outliers and finds a flatter trend, and it therefore more representative of a typical system.
Works such as \cite{Molendi2016}, \cite{Lovisari2019}, and \cite{Ghiz2020} find a $Z(R)$ over various cluster types and outer radii.
However, we note that our systems are not separated by type (i.e., cool core vs non cool core \citep{Mernier2017}, relaxed vs disturbed \citep{Lovisari2019}), and we do not isolate the Fe K$\alpha$ line as done in \cite{Ghiz2020}, instead relying on broad-band spectral fitting.

\section{Determining Metallicity Contributions to $Z_{ICM}$}
\label{sec:fitting}

\begin{figure*}[h!]
    \centering
    \includegraphics[width=1\linewidth]{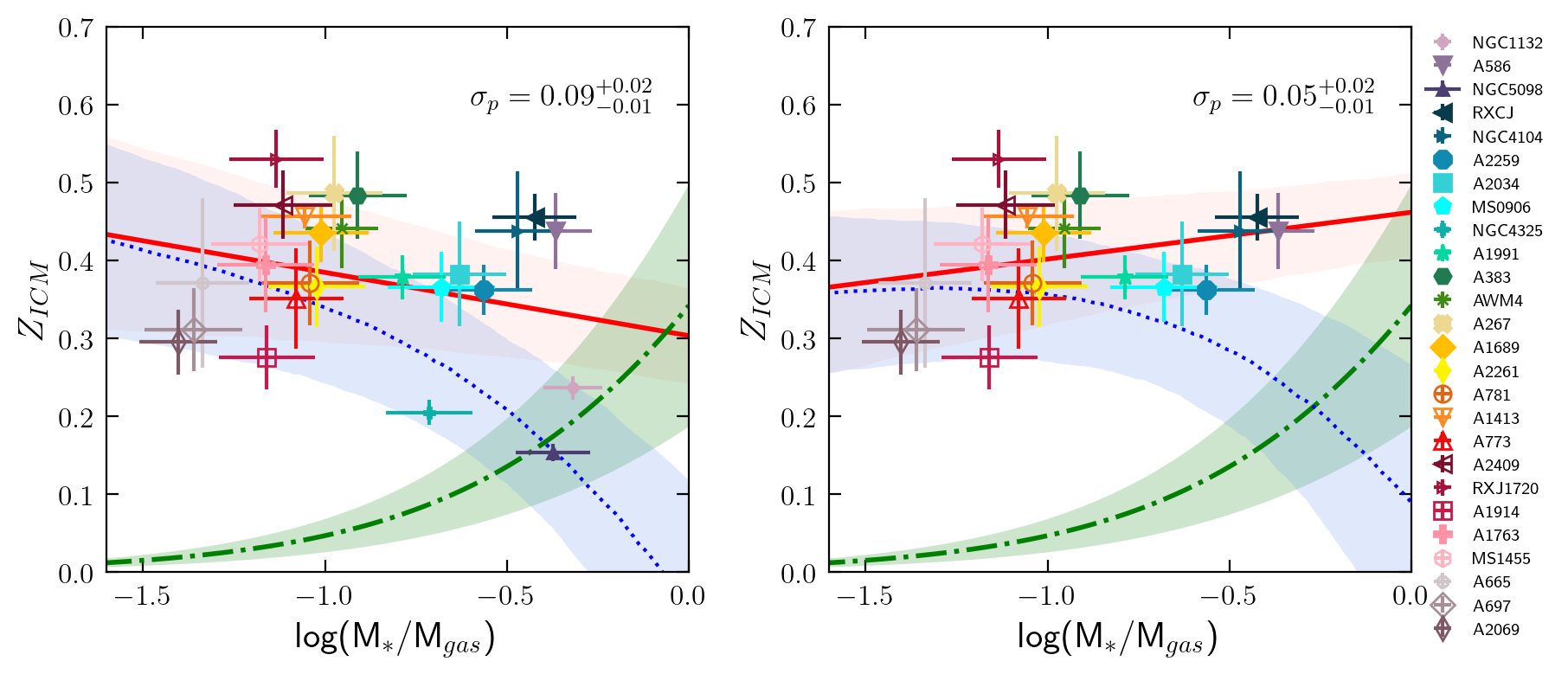}
    \caption{\textbf{Left:} We present the fit relationship of $Z_{ICM}$ vs $M_*/M_{gas}$ for all 26 groups and clusters. We find $Z_{ICM} = -0.08^{+0.07}_{-0.07}\, log(\frac{M_*}{M_{gas}}) + 0.30^{+0.06}_{-0.06}$ with an intrinsic scatter of $\sigma_p = 0.09^{+0.02}_{-0.01}$ (red solid line). This fit was done using techniques described in \cite{Sharma2017}. The derived trend of $Z_*$ (green dash-dot, Equation \ref{eq:z_star_yield}) is subtracted from $Z_{ICM}$ to find $Z_{EEP} = log_{10}\left[ \left(\frac{M_*}{M_{gas}}\right)^{-0.08^{+0.07}_{-0.07}} \left(1 + \frac{M_*}{M_{gas}}\right)^{1.14 \pm 0.52} \right] 
    + 0.30^{+0.06}_{-0.06}$ (blue dotted line). All shaded regions show the $1\sigma$ confidence intervals. \textbf{Right:} Total fit excluding NGC5098, NGC4325, and NGC1132 to determine a fit of $Z_{ICM}$ versus $M_*/M_{gas}$ as $Z_{ICM} = 0.06^{+0.06}_{-0.05}\, log(\frac{M_*}{M_{gas}}) + 0.46^{+0.05}_{-0.05}$ and $\sigma_p = 0.05^{+0.02}_{-0.01}$, shown as the red solid. We subtract off the derived $Z_*$ (green dash-dot line) to derive  $Z_{EEP} = log_{10}\left[ \left(\frac{M_*}{M_{gas}}\right)^{0.06^{+0.06}_{-0.05}} \left(1 + \frac{M_*}{M_{gas}}\right)^{1.14 \pm 0.52} \right]
    + 0.46^{+0.05}_{-0.05}$ (blue dotted line). The derived reduced $\chi^2$ is $\sim 0.90$ for all four fits, however the required $\sigma_p$ (printed in the top right corner of all plots) to keep the $\chi^2$ at an acceptable value decreases as the number of potential outliers excluded increases.}
    \label{fig:tot_fits}
\end{figure*}

The values of $Z_{ICM}$ are determined as the weighted average of the fit metallicity values in the outer three annuli.
The center two annuli are excluded as there is a known central peak in abundance due to the presence of the BCG \citep[][]{deGrandi2004, Mernier2017}.
We show the relationship of $Z_{ICM}$ and $M_*/M_{gas}$ in Figure \ref{fig:tot_fits}.
The data points show some scatter but follow a seemingly straight line.
We use the methodology described in Section 2.4 of \cite{Sharma2017} to fit the data with a straight line.
The values of $M_*/M_{gas}$ are fit in log-space due to their spread over an order of magnitude.
In a future work we investigate potential physical motivations for log $M_*/M_{gas}$ among cluster parameters (Gherri et al., in prep). 
This methodology also allows for consideration of both the x and y errors. 
The result of this fit is presented in Equation \ref{eq:sharma_fit}, and shown as the solid red line in the top left panel of Figure \ref{fig:tot_fits}.

\begin{equation}
    Z_{ICM} = -0.08^{+0.07}_{-0.07}\, log\left(\frac{M_*}{M_{gas}}\right) + 0.30^{+0.06}_{-0.06} 
\label{eq:sharma_fit}
\end{equation}

The original fit results in a reduced $\chi^2 = 5.14$.
We introduce a term for intrinsic scatter, $\sigma_p$ in the fit in order to obtain a reasonable fit. 
The factor $\sigma_p$ is determined to be $0.09^{+0.02}_{-0.01}$, resulting in a reduced $\chi^2 = 0.91$.

We also exclude the three most anomalous systems, NGC1132, NGC5098, and NGC4325 and refit for $Z_{ICM}$ finding,
\begin{equation}
    Z_{ICM} = 0.06^{+0.06}_{-0.05}\, log\left(\frac{M_*}{M_{gas}}\right) + 0.46^{+0.05}_{-0.05}.
\label{eq:sharma_fit_excl}
\end{equation}
The fit returns a reduced $\chi^2 = 2.58$ without an intrinsic scatter.
When including intrinsic scatter in the fit we obtain $\sigma_p = 0.05^{+0.02}_{-0.01}$ resulting in a reduced $\chi^2 = 0.93$.
This fit is presented as the bottom right panel in Figure \ref{fig:tot_fits}.
The other two panels in Figure \ref{fig:tot_fits} show the intermediate fits, excluding a single cluster at a time.
Further discussion on the nature of these systems is presented in Section \ref{sec:outliers}.

\subsection{Determining $Z_*$}

We first must determine the contribution of the visible stellar populations to $Z_{ICM}$, $Z_*$, in order to isolate the contribution of the EEP to $Z_{ICM}$, $Z_{EEP}$, as $Z_{EEP} = Z_{ICM} - Z_*$.
The first assumptions we make to determine $Z_*$ are that we are working with a closed box model and constant star formation. 
For the most massive systems the deep gravitational potential wells are sufficiently strong that most of the baryons are retained \citep[e.g.,][and references therin]{Kravtsov2012}.
However, simulations and observations indicate departures from this assumption in lower-mass systems, which are more susceptible to baryon loss \citep{Breg2010}.
The closed-box assumption may be appropriate for the high-mass systems, but it should be interpreted with caution, as its applicability decreases for lower-mass systems.

We begin by defining metallicity as the mass of metals, $M_z$, over the gas mass, $M_{gas}$, $Z = \frac{M_z}{M_{gas}}$.
Therefore, the change in metallicity over some time step is given by Equation \ref{eq:met_change}.
\begin{equation}
    \text{d} Z = \text{d} \Big(\frac{M_z}{M_{gas}}\Big)
\label{eq:met_change}
\end{equation}
Differentiating this, we get
\begin{equation}
    \text{d} Z =  \frac{1}{M_{gas}}(\text{d} M_z - \frac{M_z}{M_{gas}}\text{d} M_{gas}).
\label{eq:z_diff}
\end{equation}

We must now introduce more assumptions about the metal production and incorporation. 
The metal mass initially locked up in stars, $M_*$, decreases at the same rate $M_{gas}$ increases during a single time step.
That is, $-\text{d} M_* = \text{d} M_{gas}$.
For this to be correct, we make the assumption of instantaneous mixing of metals with the ICM. 
Finally, the mass of metals, $\text{d}M_z$, increases as the metals produced by stars, $\text{d} M_*$, multiplied by some yield, $y$.
We substitute $\text{d}M_z = y\text{d}M_* - Z \text{d}M_*$ into Equation \ref{eq:z_diff}, and use $\text{d}M_* = -\text{d} M_{gas}$ in order to define $Z$ solely in terms of $M_{gas}$.

\begin{equation}
    \text{d} Z = \frac{1}{M_{gas}}\left(-y \text{d} M_{gas} + Z \text{d} M_{gas} - Z \text{d}M_{gas}\right).
\end{equation}
This gives us the final relationship of
\begin{equation}
    \text{d} Z = -y \frac{\text{d} M_{gas}}{M_{gas}}
\label{eq:z_fin}
\end{equation}

We now integrate Equation \ref{eq:z_fin} from 0 to $t$ to get the total metallicity,
\begin{equation}
    Z(t) = -y \, log\Big(\frac{M_{gas}(t)}{M_{gas}(0)} \Big).
\label{eq:z_int}
\end{equation}
\vspace{0.1cm}

We next rewrite Equation \ref{eq:z_int} in terms of $M_*/M_{gas}$ to allow comparison with our fit trend of $Z_{ICM}$ vs $M_*/M_{gas}$.
The term $M_{gas}$ can be expressed as $M_{gas}(t) = M_{tot} - M_*(t)$.
At $t=0$ we assume the system to be composed entirely of gas meaning $M_{gas}(0) = M_{tot}$.
Equation \ref{eq:z_int} then becomes, 
\begin{equation}
    Z(t) = y \, log\left(1 + \frac{M_*}{M_{gas}} \right).
\label{eq:z_star}
\end{equation}

The remaining factor to define is the yield of Type Ia supernovae (SN Ia) and core collapse supernovae (SNCC), $y$.
This was thoroughly discussed and derived in \cite{Loew2013} who considered factors such as supernova rate, mass return fraction, and metals locked in stars. 
From Equations 13 and 19, we adopt an equation for yield as
\begin{equation}
    y_i = \eta^{SN} (1 - r_*)^{-1} (1 + R_{SN})^{-1} (y_i^{cc} + R_{SN} y_i^{Ia})f_{i,\odot}
\label{eq:yield}
\end{equation}
where $\eta^{SN}$ is the specific number of supernova explosions per star formed, $r_*$ is the fraction of mass previously formed into stars and recycled back into gas, $R_{SN}=\frac{\eta^{Ia}}{\eta^{cc}}$, $y_i^{cc}$ and $y_i^{Ia}$ are the yields of the $i$th element per SNCC and SN Ia respectively, and $f_{i,\odot}$ is the solar mass fraction of the $i$th element.

We adopt $\eta^{Ia} = 3.1^{+1.1}_{-1.0} \times 10^{-3}$ from DTD fits in \cite{Freund2021}.
The value for $\eta^{cc}$ is a little more complicated as it is not explicitly measured.
Few sources state calculated values of $\eta^{cc}$ such as $0.0068$ from \cite{Madau2014}, $0.007$ from \cite{Dahlen2012}, and $0.008$ from \cite{Loew2013}.
We assume a value of $\eta^{cc} = 0.0075 \pm 0.002$ to represent the average and encapsulate the range of $\eta^{cc}$ presented in the literature.
The error was chosen as it is approximately the same percent error as that on $\eta^{Ia}$, and encapsulates all possible $\eta^{cc}$.
These values, however, do not account for the metals locked up in stars. 
We adopt Equations 20 and 21 from \cite{Loew2013} as correction factors for metals locked up in stars for SN Ia and SNCC, respectively.

\cite{Ma2025} performed a recent survey using wide-field surveys from 2016-2023 to determine the fractional rates of SN Ia and SNCC.
They found $\frac{N_*^{Ia}}{N_*^{SN}} = 33.4^{+3.7}_{-11.5}\%$ and $\frac{N_*^{cc}}{N_*^{SN}} = 69.6^{+10.2}_{-20.1}\%$.
The later value was determined by adding together the determined fractions for SNe Ibc and SNe II from \cite{Ma2025}.
We multiply these values by the previously determined $\frac{N_*^{SN}}{M_*} = \eta^{Ia} + \eta^{cc} = 0.011 \pm 0.002$ to find $\eta_*^{Ia}=(2.1 \pm 0.7)\times 10^{-3}$, and $\eta_*^{cc}=(4.8 \pm 1.5)\times 10^{-3}$.
Finally, we subtract the metals locked up in stars from the total available enrichment as given in Equations 22 and 23 in \cite{Loew2013} to find $\eta^{Ia}_{ICM} = (1.0 \pm 1.1) \times 10^{-3}$ and $\eta^{cc}_{ICM} = (2.7 \pm 1.6) \times 10^{-3}$.

The final correction factor is the mass return fraction, $r_*$. 
We use Equation 6 and the cut-off points determined in \cite{Madau2014} to determine $r_*=0.35$ for a diet-Salpeter IMF.

\subsubsection{Metals Locked in Remnants}

One additional correction to consider is the metals locked in remnants. 
Throughout this derivation we have assumed constant star formation.
This means that there will be an increasing number of remnants that have locked up enriched gas from previous stellar populations.
The remnants we consider are neutron stars (NS) and black holes (BH) for SNCC, $f^{cc}$, and white dwarfs (WD) that are not in binary systems for SN Ia, $f^{Ia}$.

\cite{Kobayashi2006} provides metal yields that incorporate remnant metal lockup.
However, their approach simplifies aspects of remnant formation and metal retention, such as fixed remnant masses as a function of progenitor mass, survival fraction of WD, SN Ia pathways, binary fraction uncertainties, and the possibility of black holes being ``born in the dark".
More generally, studies of chemical evolution \citep[e.g.,][]{Werner2008, dePlaa2017, Simionescu2019} find discrepancies when using standard yield sets, including \cite{Kobayashi2006}.
This implies that a more nuanced treatment of the counting the metal budget is required.
Additional corrections must be applied to adequately capture the range of plausible scenarios for metal lock-up in remnants.
We continue with a study of metal lockup in remnants using a Bayesian approach.

The remnant lockup correction factors can be defined as 
\begin{equation}
    f^{Ia} = \frac{N^{Ia}_{WD}}{N^{Ia}}
\label{eq:corr_fac_ia}
\end{equation}
and
\begin{equation}
    f^{cc} = \frac{N^{cc}_{NS} + N^{cc}_{BH}}{N^{cc}},
\label{eq:corr_fac_cc}
\end{equation}
where
\begin{equation}
    N^{i}_{rem} = \int^{m_{hi}}_{m_{lo}} \phi(m) \frac{M_{rem}(m)}{m}dm,
\end{equation}
and
\begin{equation}
    N^{i} = \int^{m_{hi}}_{m_{lo}} \phi(m) dm.
\end{equation}
We define $m_{hi}$ and $m_{lo}$ as the high and low mass cutoff for progenitor stars, $M_{rem}(m)$ as the relationship between the progenitor and the remnant mass, and the IMF as $\phi(m)$.

The correction factor is applied to $\eta_{ICM}$ as, 
\begin{equation}
    \eta^{SN} = \eta^{cc}_{ICM}(1 - f^{cc}) + \eta^{Ia}_{ICM}(1 - f^{Ia})
\end{equation}

There is a significant amount of uncertainty surrounding the formation pathways of SN Ia, and the remnants of BHs.
To account for this, we adopt a Bayesian framework using estimates and formation theories from the literature as priors for calculations of $N^{cc}_{BH}$ and $N^{Ia}_{WD}$.

We consider WDs that are completely destroyed in SN Ia and those that are not in binary systems for the priors on $N^{Ia}_{WD}$.
There is no lockup correction when the WD is completely destroyed in a SN Ia, providing a lower prior constraint of  $N^{Ia}_{WD} = 0$.
The upper constraint is calculated by assuming $M_{rem}(m)$ from \cite{Cummings2018}, a progenitor mass range of $3-8 M_\odot$, and a binary fraction of $46\%$ \citep{Raghavan2010}.
This results in an upper limit of $N^{Ia}_{WD} = 2.69 \times 10^{-3}$.
We note that both models for $M_{rem}(m)$ provided by \cite{Cummings2018} (Equations 3 and 8) result in values of $N^{Ia}_{WD}$ within 4\%.

We must consider formation theories to determine the priors for BHs.
The first of the two competing BH formation theories is that they leave behind a remnant \citep{Blundell2008, Balakrishnan2023}.
The second is they are ``born in the dark" and are not associated with a remnant. 
A BH ``born in the dark" would result in no emission of stellar material into the ICM, resulting in an upper limit of $f^{cc} = 1$, and therefore $N^{cc}_{BH} = 8.748 \times 10^{-3}$.
We assume a progenitor mass range for BHs of $25 - 150 M_\odot$ when considering that a BH leaves behind a remnant.
The lower limit of $25 M_\odot$ follows from adopting a progenitor mass range for NS of $8-25 M_\odot$ \cite{Fryer1999, Belczynski2008}.
The upper limit is that of the diet-Salpeter IMF. 
We adopt $M_{rem}(m) = 10 M_\odot$ \citep{Heger2002}, resulting in a lower limit of $N^{cc}_{BH} = 2.17 \times 10^{-3}$.

We do not pose any priors on $N^{cc}_{NS}$ and simply adopt a progenitor mass range of $8-25 M_\odot$, and $M_{rem} (m) = 1.4 M_\odot$, calculating $N^{cc}_{Ia} = 4.19 \times 10^{-3}$. 

\begin{figure}[h!]
    \centering
    \includegraphics[width=1.0\linewidth]{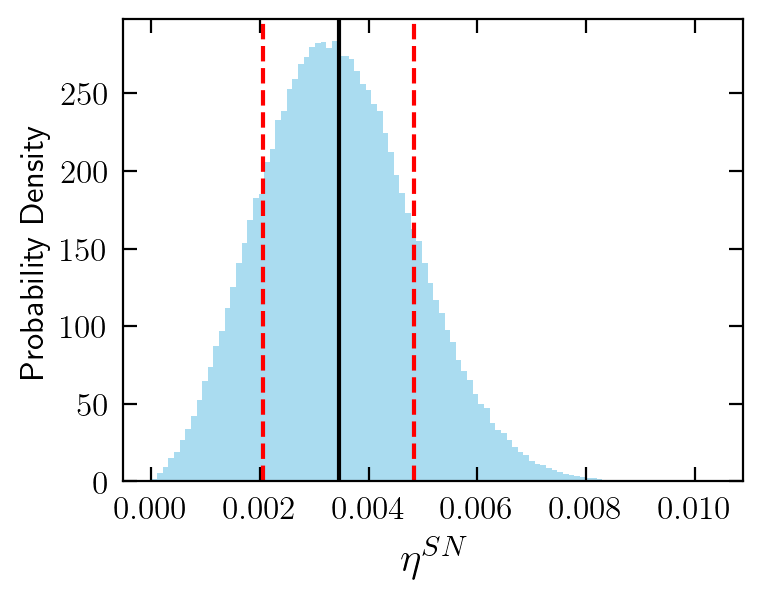}
    \caption{Posterior distribution of $\eta^{SN}$. The black vertical line shows the mean value of $0.0035$, with 1$\sigma$ shown by the dashed red lines.}
    \label{fig:eta_post}
\end{figure}

Finally, we impose a physical lower limit of 0 to the determination of $\eta^{SN}$.
These priors and constraints result in $\eta^{SN} = 3.5^{+4.8}_{-2.1} \times 10^{-3}$.
The posterior distribution is shown in Figure \ref{fig:eta_post}.

The final trend for $Z_*$ is given as Equation \ref{eq:z_star_yield}.
\begin{equation}
    Z_* = (1.14 \pm 0.52) \, log\Big(1 + \frac{M_*}{M_{gas}} \Big)
\label{eq:z_star_yield}
\end{equation}

\subsection{$Z_{EEP}$}

The trend of $Z_{EEP}$ is calculated by subtracting the expected contribution from the visible stellar populations, $Z_*$, from the fit trend of $Z_{ICM}$.
Including the entire data set, we find
\begin{equation}
\begin{split}
    Z_{EEP} = log_{10}\left[ \left(\frac{M_*}{M_{gas}}\right)^{-0.08^{+0.07}_{-0.07}} \left(1 + \frac{M_*}{M_{gas}}\right)^{1.14 \pm 0.52} \right] \\
    + 0.30^{+0.06}_{-0.06}
\label{eq:zeep_all_data}
\end{split}
\end{equation}
Excluding the outliers NGC1132 and NGC4325, $Z_{EEP}$, we determine
\begin{equation}
\begin{split}
    Z_{EEP} = log_{10}\left[ \left(\frac{M_*}{M_{gas}}\right)^{0.06^{+0.06}_{-0.05}} \left(1 + \frac{M_*}{M_{gas}}\right)^{1.14 \pm 0.52} \right] \\
    + 0.46^{+0.05}_{-0.05}
\label{eq:zeep_excl_data}
\end{split}
\end{equation}

The top left and bottom right panels of Figure \ref{fig:tot_fits} show all three derived components.

\section{Discussion}
\label{sec:discussion}

Our uniform analysis of 26 galaxy groups and clusters points to the necessity of including an EEP when considering the contributions to $Z_{ICM}$. 
Conventional closed-box chemical evolution models assume all metals are produced and retained by the visible stellar populations with more modern-day IMFs.
These classical models fail to reproduce the observed trends of $Z_{ICM}$ vs $M_*/M_{gas}$, and systematically under predict $Z_{ICM}$, especially in systems with low $M_*/M_{gas}$.
This discrepancy strongly suggests that an additional population of stars, now absent or undetectable, must have contributed significantly to the ICM metal budget.

The observed trend of $M_*/M_{gas}$ serves as a powerful constraint on the nature and origin of the EEP.
It implies that these stars must have formed early, likely pre-reionization (\citetalias{Blackwell2025}), and efficiently enriched the gas. 
As such, the trend can be used as a diagnostic to infer key properties of the EEP, such as its IMF, time of formation, and SN Ia rate as a function of redshift as done in \citetalias{Blackwell2025}.

We continue to explore the implications of several key assumptions in this model framework and discuss the influence of outlier systems on the inferred enrichment trends.

\subsection{$Z_*$ Assumptions}

The trend of $Z_{EEP}$ is inherently dependent on the derived $Z_*$.
This dependence makes our understanding of the present-day stellar populations within clusters and groups crucial.
Any uncertainty or systematic bias in $Z_*$ from observational limits, stellar population synthesis models, or assumptions of chemical yields and evolution directly propagates to $Z_{EEP}$.

Few works have performed a full analytic derivation of $Z_*$ \citep[e.g.,][]{Loew2013, Blackwell2025}.
We adopt the derivation for $y$ from \cite{Loew2013}, but adopt more modern values and add additional consideration for metals locked in remnants.
Therefore, our derived $y$ is similar to, but lower than $y = 1.55$ as estimated in \cite{Loew2013}.
\citetalias{Blackwell2025} derived a line fit for $Z_*$ assuming that the highest stellar fraction system, NGC1132, would have $Z_{ICM} = Z_*$ finding $Z_*=0.63 \pm 0.12 (M_*/M_{gas})$.

\begin{figure}[h]
    \centering
    \includegraphics[width=1\linewidth]{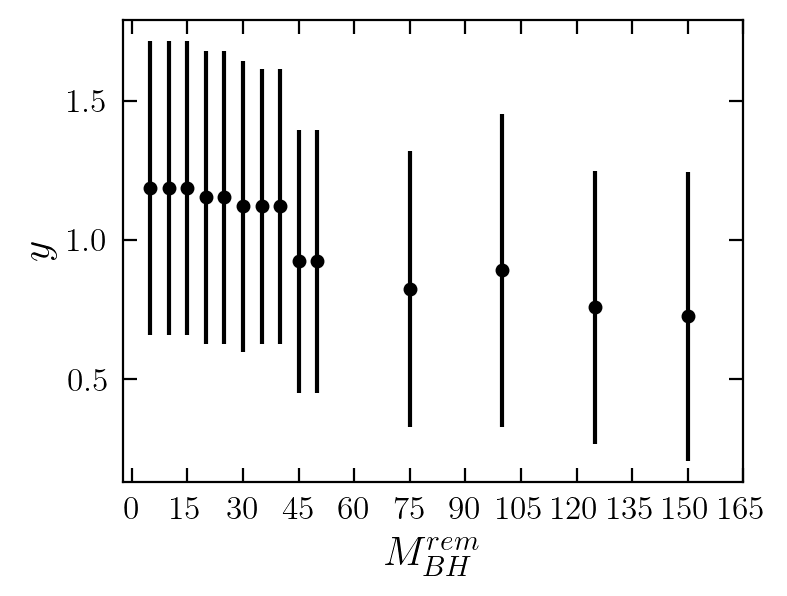}
    \caption{Relationship between the derived $y$, the yield from SNe when accounting for metals locked up in stars and remnants, as a function of assumed mean BH remnant mass, $M_{BH}^{rem}$. All resulting $y$ values are within $1\sigma$ indicating only a weak dependence on $M_{BH}^{rem}$.}
    \label{fig:y_mbh}
\end{figure}

One of the primary assumptions in the calculation of $y$ (the slope of $Z_*$) is the formation pathway of BHs. 
For the upper limit, we use the assumption that BHs are ``born in the dark".
That is, all of the progenitor stellar material goes into the formation of the black hole and no remnant is left behind.
For the lower limit, we assume that a remnant is left behind when a BH is formed and that $M_{rem}^{BH} = 10 M_\odot$ \cite{Heger2002}.
We tested the impact of this assumption on the derivation of $y$ by stepping through assumed values of $M_{rem}^{BH}$ from 5-150$M_\odot$ in steps of 5$M_\odot$ from 5-45$M_\odot$ and steps of 25$M_\odot$ from 50-150 $M_\odot$.
Figure \ref{fig:y_mbh} shows the resulting $y$ values.
We find that the assumption of $M^{rem}_{BH}$ does not change our trend of $Z_*$, as $y$ is determined within 1$\sigma$ for the full range of $M^{rem}_{BH}$.

We consider the result of this in relation of the ``born in the dark" theory.
Assuming $M_{BH}^{rem} = 150 M_\odot$ implies all progenitor stellar mass for the highest mass stars permitted by our assumed IMF collapses into the BH.
That is, there is no contribution of BH progenitors to $Z_{ICM}$. 
This results in $y = 0.724 \pm 0.519$, within $1\sigma$ of $y = 1.19\pm 0.53$ derived with $M_{BH}^{rem}=5 M_\odot$.
This lack of dependence of the yield on $M_{BH}^{rem}$ may indicate that BHs and their associated nebula do not contribute significantly to $Z_{ICM}$.
We must also consider that this trend, or lack thereof, makes sense in the context of Fe as it dominates our measurements of $Z_{ICM}$.
The biproducts of SNCC are mainly elements $\alpha$, which makes Fe a poor tracer of the dependence of $y$ on $M_\odot$.
Future work using measurements of $\alpha$ elements to trace the contribution of SNCC would be necessary to make any general conclusions using this framework.

\begin{figure}
    \centering
    \includegraphics[width=1.0\linewidth]{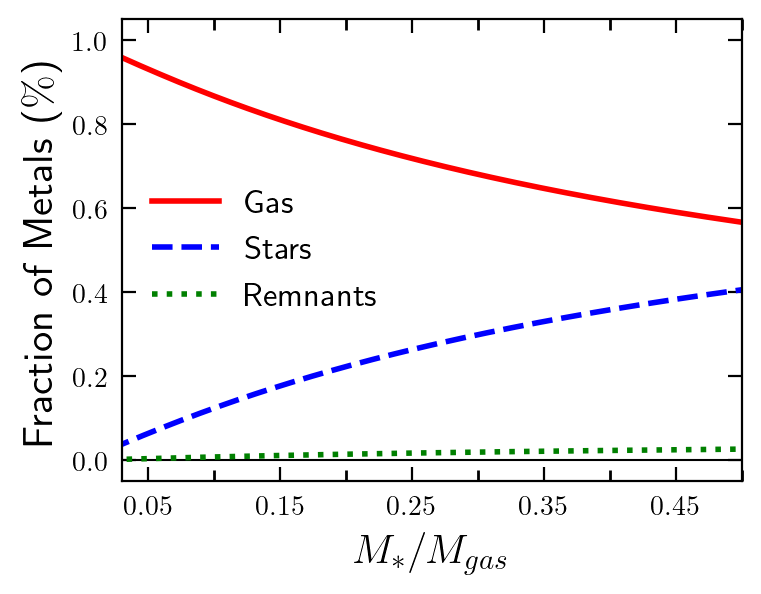}
    \caption{Fraction of metals the in gas phase (red line), locked in stars (blue dashed line), and remnants (green dotted line) as a function of $M_*/M_{gas}$.}
    \label{fig:metal_comp}
\end{figure}

We also use the derived trend of $Z_*$, and assumptions made within, to determine the partitioning of metals in the gas, stellar, and remnant phases as a function of $M_*/M_{gas}$, Figure \ref{fig:metal_comp}.
We find that the majority of metals reside in the hot gas phase for systems with low stellar mass fractions. 
As $M_*/M_{gas}$ increases, a progressively larger fraction is locked in stars. 
However, the fraction of metals locked in remnants remains nearly constant across the sample.

These findings are supported by observations and theoretical models in that a majority of metals are retained in the ICM.
Findings from \cite{Ghiz2020} support our conclusions as they find that the iron share ($M_{Fe,500}/M^*_{Fe,500}$) is approximately a factor of 10.
However, they note that large systematic uncertainties prevents them from drawing firm conclusions. 
\cite{Siv2009} shows that the BCG and ICL produce and contribute $22-31\%$ of metals in the ICM, a fraction broadly comparable to the metals fraction locked in stars.
The relatively minor metal fraction in remnants agrees with expectations of stellar evolution and feedback models \cite[e.g.,][and references therein]{Nomoto2013}.

\begin{figure*}[h!]
    \centering
    \includegraphics[width=1\linewidth]{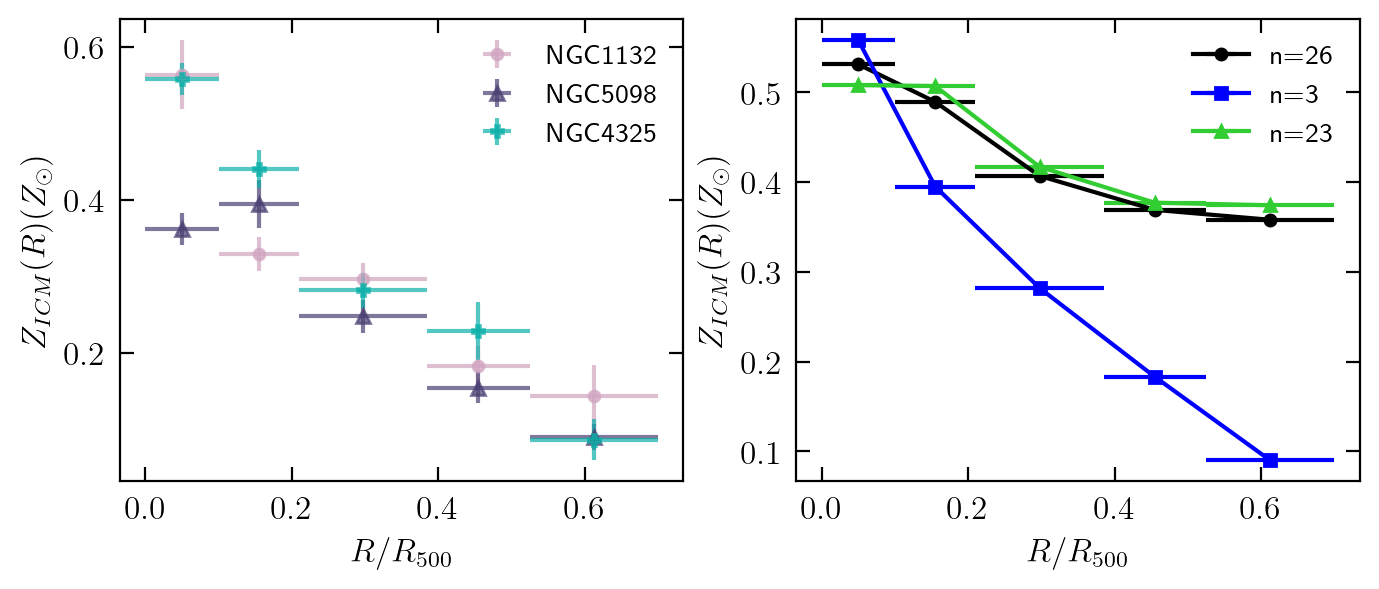}
    \caption{\textbf{Left:} Radial metallicity trends of NGC1132 (pink circles), NGC5098 (purple triangles), and NGC4525 (blue plus). Each trend is a tightly constrained and steadily decreasing with a powerlaw slope of approximately $R^{-0.5}$. \textbf{Right:} Comparison of median radial metallicity trends for samples including all 26 systems (n=26, black), just NGC1132, NGC5098, and NGC4325 (n=3, blue), and the full sample excluding NGC1132, NGC5098, and NGC4325 (n=23, green). Excluding the three anomalous systems from the full sample, n=23, results in a much flatter $Z_{ICM}$ distribution with an average $Z_{ICM}=0.37 Z_\odot$ in the outer radii. This trend is in good agreement with that of the full data set, n=26. This highlights how unique NGC1132, NGC5098, and NGC4325 are from the rest of the sample.}
    \label{fig:comp_z_rad}
\end{figure*}

\subsection{NGC1132, NGC4325, and NGC5098}
\label{sec:outliers}

Three systems in our sample, NGC1132, NGC5098, and NGC4325, stand out as potential outliers.
All systems have a high $M_*/M_{gas}$ and low $Z_{ICM}$, as summarized and presented in Table \ref{tab:total_dataset}.
Determining whether these systems are genuine outliers requires careful consideration of the adopted fit and $\sigma_p$ and the physical characteristics of each system. 

We first consider the full set of 26 groups and clusters. 
Relative to the best-fit model, NGC1132, NGC5098, and NGC4325 lie $-0.22\sigma$, $-1\sigma$ and $-0.72\sigma$ away, respectively.
However, an intrinsic scatter of $0.09^{+0.02}_{-0.01}$ required to obtain an acceptable fit statistic.
These systems also have considerable influence on the overall fit, altering the slope, intercept, and required intrinsic scatter by $~\sim 2\sigma$ when they are included vs excluded from the sample.

We consider a second case to derive a value of $\sigma_p$ that may be more representative of the survey and determine if the anomalous groups are statistical outliers.
For this we look at 16 systems with $M_*/M_{gas}$ between $0.040–0.12$ clustered with a mean $Z_{ICM}=0.40 Z_\odot$.
We assume a fit to this clump with $m=0$ and $b=0.40$ and find $\sigma_p = 0.04$ is required to achieve a fit within the 95\% confidence level.
We find that NGC1132, NGC5098, and NGC4325 are $-3.8\sigma$, $-5.9\sigma$, and $-4.5\sigma$, repspectively.
The three anomalous systems are certainly outliers in this fit scenario.
We also compare the three systems to the fit of all 26 systems  ($m=-0.08 \pm 0.07$, $b=0.30 \pm 0.06$), adopting the determined intrinsic scatter of $0.04$.
These systems fall within $-0.46\sigma$, $-2.21\sigma$, and $-1.76\sigma$ for NGC1132, NGC5098, and NGC4325, respectively.
In this case, only NGC5098 lies $>2\sigma$ away from the line of best fit and may be considered an outlier.

Based solely on statistics, we only consider NGC5098 to be an outlier as it consistently lies $>2\sigma$ away from the line of best fit.
We determine NGC1132 and NGC4325 to be anomalous, but not full outliers, since they are within $\sim 1.5 \sigma$ of the derived trends. 
While deviations of this magnitude indicate that these systems lie toward the extremes of the distribution, a $1.5\sigma$ offset corresponds to 
A confidence level of roughly 86.6\% indicates that these systems lie toward the extreme and does not strongly reject the null hypothesis that they belong to the same initial population. 
Statistical significance alone is insufficient for definitive outlier classification, making it essential to complement this analysis with a study of the physical properties and environmental context of these systems. 
This approach ensures that natural variations and measurement uncertainties are properly accounted for before classifying these systems as true outliers.

A key distinguishing feature of each group is their radial metallicity trend, shown in the left panel of Figure \ref{fig:comp_z_rad}.
NGC1132, NGC5098, and NGC4325 all exhibit steadily declining $Z_{ICM}(R)$ profiles with particularly small measurement uncertainties.
This is in contradiction to the radial trends of all 26 systems (n=26) and the sample excluding the three anomalous systems (n=23) which flatten at high radii to $0.37 Z_\odot$.
This suggests that the trend of the three anomalous systems are not representative of the general group and cluster sample.

Cosmological simulations predict varied metallicity gradients depending on the epoch and efficiency of enrichment.
Simulations by \cite{Tornatore2007} and \cite{Rasia2015} find decreasing radial metallicity trends, with steeper gradients in cool-core (CC) clusters.
In contrast, simulations by \cite{Biffi2017} find a flat radial metallicity trend with a mean $Z_{ICM} = 0.28 \pm 0.03 Z_{Fe,\odot}$.
This work uses the same metal enrichment scheme as \cite{Tornatore2007}, but with the key difference of early ($z > 2-3$) AGN feedback, allowing metal mixing with ICM.
These studies demonstrate that the slope of $Z_{ICM}(R)$ depends on the timing of star formation and the strength and epoch of AGN feedback. 
Systems that form earlier with AGN feedback display a flatter $Z_{ICM}$ distribution, while later forming groups with less AGN activity tend to centrally peaked and steeper $Z_{ICM}$ profiles.
The declining trends in our three anomalous systems qualitatively resemble predictions for groups that experienced relatively late enrichment and mixing, after reionization and contribution from the EEP.

The individual properties of these systems also support the interpretation that they may have followed atypical assembly histories and separate them from the rest of the survey. 
NGC1132 is classified as a fossil group, hosts a cool core, and has a relatively low halo mass $M_{gas,500} = (1.24\pm 0.03) \times 10^{12} M_\odot$ \cite{Mulchaey1999, Russell2007, Lagana2013, Kim2018}, consistent with efficient cooling after recent disturbance. 
NGC5098 shows disturbed kinematics, including gas sloshing, X-ray cavities correlated with radio emission, extended intragroup light, and asymmetric temperature and metallicity distributions \citep{Randall2009, Neto2025}, all indicative of recent formation and ongoing mixing. 
NGC4325, although also a cool core group, contains a Fe-rich bar-shaped structure in its center \citep{Lagana2015}, suggesting incomplete mixing and late-time enrichment.
\cite{Lagana2015} also determined the kT and Fe radial trends which are in good agreement with ours when considering that we have additional radii $>150^{\prime\prime}$ that extend to larger radii.
Together, these kinematic anomalies set these groups apart from the rest of our survey and are consistent with systems with more recent formation histories.

Altogether this suggests that NGC1132, NGC5098, and NGC4325 are outliers.
We posit that these groups formed after the peak enrichment of the EEP and post-reionization.
In this scenario, the current stellar populations of these groups would enrich the ICM, meaning their $Z_{ICM} \propto Z_*$, as seen in Figure \ref{fig:tot_fits}.
Present day stellar enrichment would also result in declining $Z_{ICM}(R)$ and low $Z_{ICM}$ compared to systems that may have had a period of early AGN feedback.
Positioning these systems in a broader theoretical context highlights the necessity of cosmological simulations for linking metallicity gradients to assembly histories \citep[e.g.,][]{Tornatore2007, Rasia2015, Biffi2017, Biffi2018}.
This has motivated further investigations into whether other distinct populations exist in the $Z_{ICM}$ vs $M_*/M_{gas}$ phase-space, and is the focus of a future work (Gherri et al., in prep).

\subsection{Revisiting the Closed-box Assumption}

A closed-box model is often assumed for studies of chemical evolution of galaxies and clusters \citep[e.g.,]{Pipino2002, Loew2013, Kudritzki2015M, Wu2015}.
In this model, all of the gas and metals are retained in the gravitational potential well.
This is a reasonable assumption for galaxy clusters as they are able to retain both primordial gas and processed metals \citep{Krav2012}.
Thus we expect the observed baryon fraction to closely trace the cosmic mean value and $Z_{ICM}$ to trace the integrated stellar yield \citep{Ettori2009, Pratt2019}.

Galaxy groups have much shallower gravitational potential wells and processes such as AGN-driven outflows may result in the loss of baryons such as hot gas and metals \citep{McCarthy2010, Young2011, Lovisari2015, Paul2017, Eckert2021, Roper2025}.
We see this in our sample resulting in lower baryon fractions for the groups, $f_{b,g} = 0.127 \pm 0.011$ than clusters, $f_{b,c} = 0.156 \pm 0.003$.
This loss of mass in groups may bias yield estimates as groups lose a greater fraction of metals per stellar mass formed, artificially steepening the relationship of $Z_{ICM}$ vs $log\left(M_*/M_{gas}\right)$.

\begin{figure}
    \centering
    \includegraphics[width=1\linewidth]{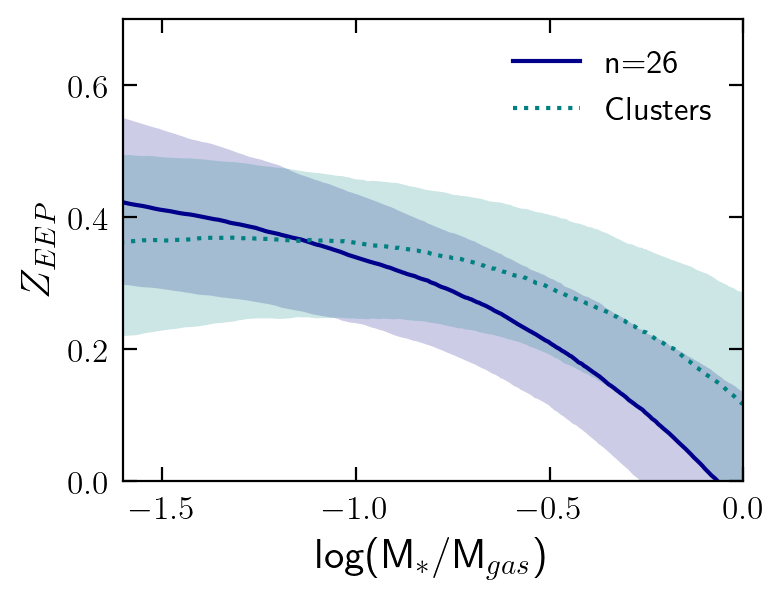}
    \caption{We compare the derived trends of $Z_{EEP}$ for the full sample (n=26, blue solid line) and just the galaxy clusters (dotted teal line). The shaded regions represent the $1\sigma$ region. The mean trend of $Z_{EEP}$ for both samples is with in $1\sigma$ of the other.}
    \label{fig:zeep_comp}
\end{figure}

We tested these systematic differences by removing the nine galaxy groups from our sample and re-deriving $Z_{EEP}$, Equation \ref{eq:zeep_clusters}.
\begin{equation}
\begin{split}
Z_{EEP} = log_{10}\left[ \left(\frac{M_*}{M_{gas}}\right)^{0.05\pm0.07} \left(1 + \frac{M_*}{M_{gas}}\right)^{1.14 \pm 0.52} \right] \\
    + 0.45^{+0.08}_{-0.07}
\label{eq:zeep_clusters}
\end{split}
\end{equation}
The intrinsic scatter determined for this fit is $\sigma_p = 0.06^{+0.02}_{-0.01}$.
These results remain consistent within $1\sigma$ when including or excluding galaxy groups (Figure \ref{fig:zeep_comp}).

The closed-box limitations for galaxy groups and variations in $f_b$ must be acknowledged and carefully considered.
However, the derived yields and primary conclusions for our cluster sample are not significantly affected by the properties of galaxy groups within our mass and redshift range.

\subsection{IMFs}

We next consider the implications of the assumed diet-Salpeter IMF.
The diet-Salpeter IMF is a modified version of the traditional Salpeter IMF. 
This IMF introduces a break in the power-law slope at $1 M_\odot$, with a flattening to $\alpha = -0.35$ in log(m) for stellar masses below this threshold. 
It also imposes a lower mass cutoff at $0.5 M_\odot$. 
Above $1 M_\odot$, the diet-Salpeter IMF is identical to the standard Salpeter IMF. 
The IMF primarily influences the determination of the stellar mass return fraction, $r*$, and the core-collapse remnant fraction, $f^{cc}$. 
Because the high-mass end of the IMF remains unchanged, adopting the original Salpeter IMF does not alter our results. 
Additionally, deviations from other commonly used IMFs, such as Kroupa or Chabrier, are minor in this context as all derived quantities are well within $1\sigma$ \citep{Kroupa2001, Chabrier2003}.

\section{Summary}
\label{sec:conclusions}

The presence of an EEP has broad implications for the history of star formation in the early Universe and the interpretation of chemical evolution in large-scale structures. 
The EEP may reflect a period of intense star formation of late Pop III and early Pop II stars potentially linked to early feedback processes that shaped the abundance of the ICM \citep{Elbaz1995, Loew2013, Renzini2014, Blackwell2025}. 

There are limited constraints on the EEP, such as those in \cite{Loew2013} and \citetalias{Blackwell2025}.
This is due to the reliance on uniform $Z_{ICM}$ measurements and robust estimates of the metals contributed from the present day stellar populations $Z_*$.
The goal of this survey was to perform a uniform analysis of 26 galaxy groups and clusters, with consistent measurements of $M_{*,500}$ and $M_{gas,500}$ to minimize uncertainties on $Z_{ICM}$, and therefore the metal contribution from the EEP $Z_{EEP}$.
Here we summarize our key findings from this survey work: 

\begin{enumerate}
    \item We used an MCMC and fitting statistics defined in \cite{Sharma2017} to fit the full sample of clusters to find Equation \ref{eq:sharma_fit} with an intrinsic scatter of $\sigma_p = 0.09^{+0.02}_{-0.01}$ with a reduced $\chi^2 = 0.91$. Excluding likely outliers NGC1132, NGC5098, and NGC4325 we fit Equation \ref{eq:sharma_fit_excl} with an intrinsic scatter of $\sigma_p = 0.05^{+0.05}_{-0.01}$ with a reduced $\chi^2=0.93$.

    \item We present a detailed look at the derivation of $Z_*$, demonstrating that the visible stellar populations alone systematically underpredicts the observed $Z_{ICM}$, particularly in systems with low $M_*/M_{gas}$. For this derivation, we begin with the assumption of a closed-box system, continuous star formation, and instantaneous mixing. We used the yield determination from \cite{Loew2013}, and added an additional consideration of the metals locked in remnants to derive $Z_*$ as presented in Equation \ref{eq:z_star_yield}. 

    \item We derive the inferred $Z_{EEP}$ as the difference between $Z_{ICM} - Z_*$, presented in Equations \ref{eq:zeep_all_data} and \ref{eq:zeep_excl_data}. The best-fit relations in Equations \ref{eq:sharma_fit} and \ref{eq:sharma_fit_excl} reveal a nearly flat or slightly increasing trend, reinforcing the need for an EEP.

    \item We identified three outlier systems NGC1132, NGC5098, and NGC4325. They each exhibit low $Z_{ICM}$ and pronounced negative radial $Z_{ICM}$ trends, despite their high relatively $M_*/M_{gas}$. Their inclusion in global metallicity fits significantly affects the inferred slope and intrinsic scatter ($\sim2\sigma$), highlighting their influence and uniqueness. We posit that these systems formed post-reionization after the rest of the systems within the sample and peak of the EEP. Such characteristics are consistent with cosmological simulations that show later forming groups tend to have steeper metallicty gradients and a lower $Z_{ICM}$ \citep{Tornatore2007, Rasia2015, Biffi2017}. 

\end{enumerate}

This study confirms and quantifies the necessity of a population such as the EEP to explain $Z_{ICM}$.
We provide a robust and testable framework for future theoretical and observational work on the origin and physical properties of the EEP.

\section{Acknowledgments}
We greatly thank Cameron Pratt, Zhijie Qu, Rui Hang, and Jiangtao Li for thoughtful discussions, guidance, and feedback on this work.
The authors would like to acknowledge the support from NASA ADAP Award 80NSSC22K0481 for this program. 

\bibliography{main}{}
\bibliographystyle{aasjournal}


\afterpage{
\clearpage
\begin{table}[htbp]
\rotatebox{90}{%
\hspace*{\dimexpr-0.8\columnwidth-\columnsep}\begin{minipage}{\textheight}
\centering
\caption{Total Data Set}
\label{tab:total_dataset}
\vspace{1em}
\renewcommand{\arraystretch}{1.5}
    \begin{tabular}{c | c | c | c | c | c | c | c}
        \toprule
 Cluster & RA & Dec & z & $R_{500}$ (kpc) & $M_{gas,500}$ ($M_\odot$) & $M_*/M_{gas}$ & $Z_{ICM}$ ($Z_\odot$) \\

        \hline

         NGC1132 & 02h52m51.9s & -01d16m29s & 0.023 & 383.5 & $(1.24 \pm 0.03) \times 10^{12}$ & 0.48 $\pm$ 0.09 & 0.24 $\pm$ 0.02 \\

         A586 & 07h32m20.160s & +31d37m55.92s & 0.17 & 1265.8 & $(6.9 \pm 0.06) \times 10^{13}$& 0.43 $\pm$ 0.10 & 0.44 $\pm$ 0.05 \\

         NGC5098 & 13h20m16.2s & +33d08m39s & 0.037 & 344.6 & $(1.3 \pm 0.21) \times 10^{12}$ & 0.42 $\pm$ 0.13 & 0.15 $\pm$ 0.01 \\

         RXCJ2315.7-0222 & 23h15m44.1s & -02d22m59s & 0.027 & 427.1 & $(1.86 \pm 0.19)\times 10^{12}$ & 0.38 $\pm$ 0.10 & 0.46 $\pm$ 0.03  \\

         NGC4104 & 12h06m39.0s & +28d10m27s & 0.028 & 502.5 & $(2.05 \pm 0.09) \times 10^{12}$ & 0.34 $\pm$ 0.09 & 0.44 $\pm$ 0.08  \\
         
         A2259 & 17h20m10.080s & +27d39m03.28s & 0.16 & 1117.1 & $(5.5 \pm 0.4) \times 10^{13}$ & 0.27 $\pm$ 0.08 & 0.36 $\pm$ 0.03  \\

         A2034 & 15h10m12.480s & +33d30m28.08s & 0.11 & 1315.5 & $(7.3 \pm 0.2)\times 10^{13}$ & 0.23 $\pm$ 0.07 & 0.38 $\pm$ 0.07 \\

         MS0906.5+1110 & 09h09m12.720s & +10d48m32.88s & 0.18 & 923.0 & $(4.4 \pm 0.2)\times 10^{13}$ & 0.21 $\pm$ 0.07 & 0.37 $\pm$ 0.04 \\

         NGC4325 & 12h33m06.7s & +10d37m16s & 0.026 & 408.6 & $(1.50 \pm 0.02)\times 10^{12}$ & 0.19 $\pm$ 0.05 & 0.21 $\pm$ 0.02  \\

         A1991 & 14h54m30.2s & +18d37m51s & 0.059 & 563.3 & $(6.68 \pm 0.12)\times 10^{12}$ & 0.16 $\pm$ 0.05 & 0.38 $\pm$ 0.03 \\

         A383 & 02h48m03.432s & -03d31m45.87s & 0.19 & 971.2 & $(4.0 \pm 0.2)\times 10^{13}$ & 0.12 $\pm$ 0.04 & 0.48 $\pm$ 0.06 \\

         AWM4 & 16h04m57.0s & +23d55m14s & 0.032 & 547.8 & $(5.60 \pm 0.29)\times 10^{12}$ & 0.11 $\pm$ 0.03 & 0.44 $\pm$ 0.05  \\

         A267 & 01h52m42.144s & +01d00m41.29s & 0.23 & 953.3 & $(5.2 \pm 0.5)\times 10^{13}$ & 0.11 $\pm$ 0.03 & 0.49 $\pm$ 0.07  \\

         A1689 & 13h11m29.520s & -01d20m29.68s & 0.18 & 1487.5 & $(12.0 \pm 0.5)\times 10^{13}$ & 0.098 $\pm$ 0.029 & 0.44 $\pm$ 0.04 \\

         A2261 & 17h22m27.120s & +32d07m56.28s & 0.22 & 1176.8 & $(9.7 \pm 0.4)\times 10^{13}$ & 0.095 $\pm$ 0.029 & 0.37 $\pm$ 0.05  \\

         A781 & 09h20m26.160s & +30d30m02.52s & 0.30 & 984.4 & $(6.6 \pm 0.5)\times 10^{13}$ & 0.091 $\pm$ 0.028 & 0.37 $\pm$ 0.05\\

         A1413 & 11h55m18.000s & +23d24m17.28s & 0.14 & 1330.7 & $(8.4 \pm 0.3)\times 10^{13}$ & 0.088 $\pm$ 0.026 & 0.46 $\pm$ 0.05\\

         A773 & 09h17m53.040s & +51d43m39.72s & 0.22 & 1306.3 & $(9.6 \pm 0.4)\times 10^{13}$ & 0.083 $\pm$ 0.025 & 0.35 $\pm$ 0.06  \\

         RXJ1720.1+2638 & 17h20m16.800s & +26d38m06.60s & 0.39 & 1178.4 & $(7.3 \pm 0.4)\times 10^{13}$ & 0.077 $\pm$ 0.024 & 0.53 $\pm$ 0.04  \\

          A2409 & 22h00m52.800s & +20d58m27.84s & 0.15 & 1177.1 & $(6.4 \pm 0.4) \times 10^{13}$ & 0.073 $\pm$ 0.022 & 0.47 $\pm $0.04 \\

         A1914 & 14h26m00.960s & +37m49d33.96s & 0.17 & 1503.5 & $(11.7 \pm 0.7) \times 10^{13}$ & 0.069 $\pm$ 0.021 & 0.28 $\pm $0.04 \\

         A1763 & 13h35m18.240s & +40d59m59.28s & 0.22 & 1328.3 & $(11.5 \pm 0.6) \times 10^{13}$ & 0.069 $\pm$ 0.021 & 0.39 $\pm$ 0.06 \\

         MS1455.0+2232 & 14h57m15.120s & +22d20m35.52s & 0.26 & 930.3 & $(5.0 \pm 0.2) \times 10^{13}$ & 0.066 $\pm$ 0.020 & 0.42 $\pm$ 0.05  \\

         A665 & 08h30m57.360s & +65d50m33.36s & 0.18 & 1357.3 & $(12.6 \pm 0.5) \times 10^{13}$& 0.046 $\pm$ 0.014 & 0.37 $\pm$ 0.11 \\

         A697 & 08h42m57.600s & +36d21m55.80s & 0.28 & 1455.7 & $(16.1 \pm 1.1) \times 10^{13}$ & 0.044 $\pm$ 0.013 & 0.31 $\pm$ 0.05  \\

         A2069 & 15h24m39.849s & +29d53m26.33s & 0.12 & 1173.2 & $(6.3 \pm 0.3) \times 10^{13})$ & 0.040 $\pm$ 0.010 & 0.30 $\pm $0.04  \\  
    \bottomrule
    \end{tabular}
\end{minipage}
}
\end{table}
\clearpage
}

\afterpage{
\clearpage
\begin{table}[htbp]
\rotatebox{90}{%
\hspace*{\dimexpr-0.8\columnwidth-\columnsep}\begin{minipage}{\textheight}
\centering
\caption{kT (keV)}
\label{tab:kT_summary}
\vspace{1em}
\renewcommand{\arraystretch}{1.5}
    \begin{tabular}{l l l l l l l}
        \toprule
Cluster & obs ID & $< 0.1 R_{500}$ & $0.1-0.21 R_{500}$ & $0.21 - 0.39 R_{500}$ & $0.39 - 0.53 R_{500}$ & $0.53 - 0.7 R_{500}$ \\
\hline
NGC1132 & 0151490101 & 1.184 $\pm$ 0.010 & 1.146 $\pm$ 0.094 & 1.189 $\pm$ 0.011 & 1.103 $\pm$ 0.022 & 1.047 $\pm$ 0.040 \\

A586 & 827051101 & 6.584 $\pm$ 0.185 & 11.284 $\pm$ 0.795 & 6.445 $\pm$ 0.216 & 7.694 $\pm$ 0.690 & 3.942 $\pm$ 0.450 \\
   & 0673850201 & 7.107 $\pm$ 0.264 & 6.927 $\pm$ 0.278 & 4.879 $\pm$ 0.243 & 5.837 $\pm$ 0.594 & 5.293 $\pm$ 1.439 \\

NGC5098 & 0105860101 & 1.023 $\pm$ 0.006 & 1.205 $\pm$ 0.011 & 1.233 $\pm$ 0.017 & 1.069 $\pm$ 0.019 & 0.956 $\pm$ 0.024 \\

RXCJ 2315.7-0222 & 501110101 & 1.560 $\pm$ 0.015 & 1.761 $\pm$ 0.027 & 1.651 $\pm$ 0.030 & 1.496 $\pm$ 0.040 & 1.536 $\pm$ 0.066 \\

NGC4104 & 0301900401 & 1.496 $\pm$ 0.033 & 2.334 $\pm$ 0.114 & 2.000 $\pm$ 0.106 & 1.823 $\pm$ 0.211 & 2.639 $\pm$ 0.746 \\

A2259 & 0827040101 & 5.980 $\pm$ 0.165 & 5.731 $\pm$ 0.120 & 4.963 $\pm$ 0.145 & 5.091 $\pm$ 0.191 & 4.931 $\pm$ 0.386 \\

A2034 & 149880101 & 4.772 $\pm$ 0.214 & 6.279 $\pm$ 0.240 & 6.584 $\pm$ 0.331 & 7.175 $\pm$ 0.778 & 6.435 $\pm$ 1.091 \\
   & 303930101 & 7.322$ \pm$ 0.304 & 7.399 $\pm$ 0.346 & 7.747$ \pm$ 0.453 & 4.575 $\pm$ 0.444 & 6.298 $\pm$ 1.587 \\

MS0906.5+1110 & 673850901 & 6.435 $\pm$ 0.468 & 5.635 $\pm$ 0.363 & 5.632 $\pm$ 0.249 & 5.460 $\pm$ 0.408 & \\
   & 827351301 & 6.632 $\pm$ 0.527 & 5.971 $\pm$ 0.348 & 5.328 $\pm$ 0.279 & 5.637 $\pm$ 0.230 & 5.283 $\pm$ 0.563 \\

NGC4325 & 0108860101 & 0.959 $\pm$ 0.003 & 1.099 $\pm$ 0.007 & 1.052 $\pm$ 0.010 & 0.979 $\pm$ 0.018 & 0.929 $\pm$ 0.041 \\

A1991 & 0145020101 & 1.805 $\pm$ 0.013 & 2.953 $\pm$ 0.066 & 2.558 $\pm$ 0.515 & 2.174 $\pm$ 0.088 & 2.037 $\pm$ 0.151 \\

A383 & 0084230501 & 3.463 $\pm$ 0.058 & 4.726 $\pm$ 0.117 & 4.837 $\pm$ 0.142 & 4.580 $\pm$ 0.260 & 4.588 $\pm$ 0.430 \\

AWM4 & 0093060401 & 2.490 $\pm$ 0.041 & 2.504 $\pm$ 0.049 & 2.016 $\pm$ 0.072 & 2.710 $\pm$ 0.190 & 1.849 $\pm$ 0.178 \\

A267 & 0084230401 & 6.629 $\pm$ 0.059 & 5.936 $pm$ 0.298 & 5.801 $\pm$ 0.270 & 5 & 5.109 $\pm$ 0.444 \\

A1689 & 0093030101 & 8.864$ \pm$ 0.144 & 8.490 $\pm$ 0.188 & 8.835 $\pm$ 0.019 & 9.275 $\pm$ 0.432 & \\

A2261 & 6931809101 & 7.362 $\pm$ 0.218 & 7.869 $\pm$ 0.211 & 8.156 $\pm$ 0.357 & 7.833 $\pm$ 0.414 & 6.078 $\pm$ 0.328 \\

A781 & 0401170101 & 5.913 $\pm$ 1.518 & 6.264 $\pm$ 0.558 & 5.241 $\pm$ 0.218 & 6.813 $\pm$ 0.427 & 5.592 $\pm$ 0.276 \\

A1413 & 0502690101 & 7.204 $\pm$ 0.131 & 6.728 $\pm$ 0.147 & 7.059 $\pm$ 0.159 & 6.730 $\pm$ 0.274 & 6.912 $\pm$ 0.582 \\
   & 0502690201 & 7.151 $\pm$ 0.086 & 7.450 $\pm$ 0.113 & 7.352 $\pm$ 0.115 & 7.413 $\pm$ 0.235 & 7.606 $\pm$ 0.288 \\
   & 0551280101 & 7.255 $\pm$ 0.098 & 7.100 $\pm$ 0.107 & 7.445 $\pm$ 0.128 & 7.178 $\pm$ 0.225 & 7.028 $\pm$ 0.373 \\
   & 0551280201 & 7.173 $\pm$ 0.096 & 7.483 $\pm$ 0.134 & 7.001 $\pm$ 0.117 & 7.020 $\pm$ 0.223 & 6.506 $\pm$ 0.367 \\
    \bottomrule
    \end{tabular}
\end{minipage}
}
\end{table}
\clearpage
}

\afterpage{
\clearpage
\begin{table}[htbp]
\rotatebox{90}{%
\hspace*{\dimexpr-0.8\columnwidth-\columnsep}\begin{minipage}{\textheight}
\centering
\caption{Table \ref{tab:kT_summary} cot.}
\label{tab:kT_summary2}
\vspace{1em}
\renewcommand{\arraystretch}{1.5}
    \begin{tabular}{l l l l l l l}
        \toprule
Cluster & obs ID & $< 0.1 R_{500}$ & $0.1-0.21 R_{500}$ & $0.21 - 0.39 R_{500}$ & $0.39 - 0.53 R_{500}$ & $0.53 - 0.7 R_{500}$ \\
\hline
A773 & 0084230601 & 9.694 $\pm$ 0.869 & 7.865 $\pm$ 0.298 & 7.797 $\pm$ 0.270 & 7.352 $\pm$ 0.524 & 8.256 $\pm$ 0.833 \\

RXJ1720.1+2638 & 0500670201 & 4.194 $\pm$ 0.095 & 6.763 $\pm$ 0.165 & 6.760 $\pm$ 0.275 & 8.119 $\pm$ 0.612 & 6.5 \\
   & 0500670301 & 4.908 $\pm$ 0.065 & 6.727 $\pm$ 0.149 & 7.389 $\pm$ 0.229 & 6.670 $\pm$ 0.354 & 6.640 $\pm$ 0.565 \\
   & 0500670401 & 5.018 $\pm$ 0.081 & 6.864 $\pm$ 0.202 & 6.652 $\pm$ 0.237 & 6.893 $\pm$ 0.472 & 6.5 \\

A2409 & 0827020101 & 9.694 $\pm$ 0.869 & 5.888 $\pm$ 0.156 & 6.100 $\pm$ 0.151 & 5.551 $\pm$ 0.236 & 6.541 $\pm$ 0.555 \\

A1914 & 0112230201 & 6.664 $\pm$ 0.159 & 5.586 $\pm$ 0.149 & 6.209 $\pm$ 0.234 & 7.465 $\pm$ 0.440 & 7.292 $\pm$ 0.610 \\

A1763 & 0084230901 & 6.196 $\pm$ 0.473 & 6.040 $\pm$ 0.306 & 5.660 $\pm$ 0.237 & 5.582 $\pm$ 0.354 & 4.532 $\pm$ 0.284 \\

MS1455+2232 & 0108670201 & 3.959 $\pm$ 0.063 & 4.717 $\pm$ 0.092 & 5.050 $\pm$ 0.133 & 5.385 $\pm$ 0.246 & 4.420 $\pm$ 0.285 \\

A665 & 0109890401 & 6.938 $\pm$ 0.721 & 6.949 $\pm$ 0.551 & 7.048 $\pm$ 0.514 & 7.396 $\pm$ 0.814 & 6.801 $\pm$ 0.957 \\

A697 & 0827041001 & 10.334 $\pm$ 0.700 & 8.787 $\pm$ 0.409 & 9.938 $\pm$ 0.357 & 9.233 $\pm$ 0.522 & 8.388 $\pm$ 0.644 \\

A2069 & 0827010601 & 6.178 $\pm$ 0.323 & 7.023 $\pm$ 0.259 & 5.993 $\pm$ 0.160 & 9.008 $\pm$ 0.473 & 6.194 $\pm$ 0.451 \\
 
    \bottomrule
    \end{tabular}
\end{minipage}
}
\end{table}
\clearpage
}

\afterpage{
\clearpage
\begin{table}[htbp]
\rotatebox{90}{%
\hspace*{\dimexpr-0.8\columnwidth-\columnsep}\begin{minipage}{\textheight}
\centering
\caption{$Z_{ICM} (Z_\odot)$}
\label{tab:z_summary}
\vspace{1em}
\renewcommand{\arraystretch}{1.5}
    \begin{tabular}{l l l l l l l}
        \toprule
Cluster & obs ID & $< 0.1 R_{500}$ & $0.1-0.21 R_{500}$ & $0.21 - 0.39 R_{500}$ & $0.39 - 0.53 R_{500}$ & $0.53 - 0.7 R_{500}$ \\
\hline

NGC1132 & 0151490101 &  0.564 $\pm$ 0.045 & 0.330 $\pm$ 0.022 & 0.297 $\pm$ 0.021 & 0.183 $\pm$ 0.026 & 0.144 $\pm$ 0.040 \\

A586 & 827051101 & 0.620 $\pm $0.063 & 0.235 $\pm$ 0.129 & 0.534 $\pm$ 0.076 & 0.600 $\pm$ 0.196 & 0.127 $\pm $0.159 \\
    & 0673850201 & 0.524 $\pm$ 0.083 & 0.421 $\pm$ 0.078 & 0.411 $\pm$0.079 & 0.231 $\pm$  0.213 & 0.527 $\pm$ 0.707 \\

NGC5098 & 0105860101 & 0.362 $\pm$ 0.021 & 0.395 $\pm$ 0.031 & 0.248 $\pm$ 0.021 & 0.155 $\pm$ 0.020 & 0.090 $\pm$ 0.017 \\

RXCJ 2315.7-0222 & 501110101 & 0.791 $\pm $0.041 & 0.645 $\pm$ 0.043 & 0.495 $\pm$ 0.037 & 0.377 $\pm$ 0.050 & 1.589 $\pm$ 0.656 \\

NGC4104 & 0301900401 & 0.228 $\pm$ 0.026 & 0.458 $\pm$ 0.083 & 0.425 $\pm$ 0.083 & 0.482 $\pm$ 0.206 & 0.748 $\pm$ 0.604 \\

A2259 & 0827040101 & 0.508 $\pm$ 0.063 & 0.442 $\pm$ 0.045 & 0.364 $\pm$ 0.036 & 0.404 $\pm$ 0.085 & 0.247 $\pm$ 0.130 \\

A2034 & 149880101 & 0.360 $\pm$ 0.073 & 0.530 $\pm$ 0.076 & 0.391 $\pm$ 0.097 & 0.451 $\pm$ 0.247 & 1.786 $\pm$ 0.0896 \\
    & 303930101 &  0.400 $\pm$ 0.079 & 0.465 $\pm$ 0.072 & 0.399 $\pm$ 0.113 & 7e-4, $\pm$ 0.258 & 0.626 $\pm$ 0.602\\

MS0906.5+1110 & 673850901 & 0.753 $\pm$ 0.203 & 0.456 $\pm$ 0.119 & 0.539 $\pm$ 0.108 & 0.119 $\pm$ 0.177 & \\
    & 827351301 & 0.437 $\pm$ 0.156 & 0.546 $\pm$ 0.104 & 0.340 $\pm$ 0.075 & 0.363 $\pm$ 0.076 & 0.296 $\pm$ 0.195 \\

NGC4325 & 0108860101 & 0.558 $\pm$ 0.021 & 0.441 $\pm$ 0.025 & 0.282 $\pm$ 0.023 & 0.229 $\pm$ 0.038 & 0.087 $\pm$ 0.027\\

A1991 & 0145020101 & 0.651 $\pm$ 0.023 & 0.831 $\pm$ 0.059 & 0.472 $\pm$ 0.039 & 0.262 $\pm$ 0.045 & 0.338 $\pm$ 0.100 \\

A383 & 0084230501 & 0.857 $\pm$ 0.057 & 0.626 $\pm$ 0.065 & 0.551 $\pm$ 0.069 & 0.315 $\pm$ 0.108 & 0.495 $\pm$ 0.217 \\

AWM4 & 0093060401 & 0.740 $\pm$ 0.047 & 0.550 $\pm$ 0.041 & 0.409 $\pm$ 0.056 & 0.532 $\pm$ 0.124 & 0.862 $\pm$ 0.321\\

A267 & 0084230401 & 0.493 $\pm$ 0.192 & 0.589 $\pm$ 0.113 & 0.490 $\pm$ 0.093 & 0.451 $\pm$ 0.145 & 0.546 $\pm$ 0.222\\

A1689 & 0093030101 & 0.501 $\pm$ 0.032 & 0.358 $\pm$ 0.034 & 0.475 $\pm$ 0.042 & 0.290 $\pm$ 0.080 &\\

A2261 & 6931809101 & 0.630 $\pm$ 0.065 & 0.534 $\pm$ 0.054 & 0.293 $\pm$ 0.072 & 0.504 $\pm$ 0.104 & 0.382 $\pm$ 0.104 \\

A781 & 0401170101 & 0.508 $\pm$ 0.552 & 0.345 $\pm$ 0.164 & 0.397 $\pm$ 0.083 & 0.316 $\pm$ 0.119 & 0.372 $\pm$ 0.092\\

A1413 & 0502690101 & 0.638 $\pm$ 0.043 & 0.437 $\pm$ 0.041 & 0.472 $\pm$ 0.046 & 0.522 $\pm$ 0.090 & 0.034 $\pm$ 0.135 \\
    & 0502690201 & 0.588 $\pm$ 0.028 & 0.467 $\pm$ 0.031 & 0.427 $\pm$ 0.029 & 0.421 $\pm$ 0.060 & 0.407 $\pm$ 0.079 \\
    & 0551280101 & 0.649 $\pm$ 0.030 & 0.528 $\pm$ 0.033 & 0.527 $\pm$ 0.036 & 0.389 $\pm$ 0.063 & 0.478 $\pm$ 0.108 \\
    & 0551280201 & 0.571 $\pm$ 0.030 & 0.416 $\pm$ 0.035 & 0.481 $\pm$ 0.034 & 0.460 $\pm$ 0.066 & 0.434 $\pm$ 0.106 \\
    \bottomrule
    \end{tabular}
\end{minipage}
}
\end{table}
\clearpage
}

\afterpage{
\clearpage
\begin{table}[htbp]
\rotatebox{90}{%
\hspace*{\dimexpr-0.8\columnwidth-\columnsep}\begin{minipage}{\textheight}
\centering
\caption{Table \ref{tab:z_summary} cont.}
\label{tab:z_summary2}
\vspace{1em}
\renewcommand{\arraystretch}{1.5}
    \begin{tabular}{l l l l l l l}
        \toprule
Cluster & obs ID & $< 0.1 R_{500}$ & $0.1-0.21 R_{500}$ & $0.21 - 0.39 R_{500}$ & $0.39 - 0.53 R_{500}$ & $0.53 - 0.7 R_{500}$ \\
\hline

A773 & 0084230601 & 0.323 $\pm$ 0.159 & 0.516 $\pm$ 0.101 & 0.367 $\pm$ 0.081 & 0.307 $\pm$ 0.129 & 0.358 $\pm$ 0.183 \\

RXJ1720.1+2638 & 0500670201 & 0.797 $\pm$ 0.085 & 0.516 $\pm$ 0.065 & 0.401 $\pm$ 0.087 & 0.676 $\pm$ 0.212 & 0.418 $\pm$ 0.274 \\
    & 0500670301 & 0.734 $\pm$ 0.036 & 0.557 $\pm$ 0.049 & 0.470 $\pm$ 0.067 & 0.512 $\pm$ 0.114 & 0.589 $\pm$ 0.193 \\
    & 0500670401 & 0.763 $\pm$ 0.047 & 0.572 $\pm$ 0.065 & 0.712 $\pm$ 0.089 & 0.593 $\pm$ 0.159 & 0.834 $\pm$ 0.239\\

A2409 & 0827020101 & 0.323 $\pm$ 0.159 & 0.566 $\pm$ 0.061 & 0.466 $\pm$ 0.052 & 0.566 $\pm$ 0.097 & 0.284 $\pm$ 0.152 \\

A1914 & 0112230201 & 0.372 $\pm$ 0.043 & 0.204 $\pm$ 0.031 & 0.238 $\pm$ 0.046 & 0.314 $\pm$ 0.103 & 0.781 $\pm$ 0.189 \\

A1763 & 0084230901 & 0.624 $\pm$ 0.177 & 0.545 $\pm$ 0.109 & 0.367 $ \pm$ 0.079 & 0.481 $\pm$ 0.132 & 0.381 $\pm$ 0.144 \\

MS1455+2232 & 0108670201 & 0.625 $\pm$ 0.044 & 0.527 $\pm$ 0.046 & 0.428 $\pm$ 0.057 & 0.462 $\pm$ 0.103 & 0.291 $\pm$ 0.146 \\

A665 & 0109890401 & 0.188 $\pm$ 0.176 & 0.483 $\pm$ 0.154 & 0.417 $\pm$ 0.139 & 0.373 $\pm$ 0.243 & 0.220 $\pm$ 0.252 \\

A697 & 0827041001 & 0.478 $\pm$ 0.130 & 0.386 $\pm$ 0.075 & 0.359 $\pm$ 0.069 & 0.365 $\pm$ 0.110 & 0.069 $\pm $0.130 \\

A2069 & 0827010601 & 0.503 $\pm$ 0.113 & 0.419 $\pm$ 0.071 & 0.392 $\pm$ 0.053 & 0.128 $\pm$ 0.084 & 0.163 $\pm$ 0.112 \\
 
    \bottomrule
    \end{tabular}
\end{minipage}
}
\end{table}
\clearpage
}

\end{document}